\input harvmac
\noblackbox
\input epsf.tex
%
%

\font\tiau=cmcsc10
\baselineskip 12pt
\Title{\vbox{\baselineskip12pt \hbox{}
\hbox{} }}
{\vbox{\hbox{\centerline{\bf ASPECTS OF LOCALIZED GRAVITY AROUND THE SOFT MINIMA}}}}
\vskip3cm
\centerline{\tiau Pablo M. Llatas\foot{llatas@fpaxp1.usc.es}}

\vskip.1in
\centerline{\it Departamento de F\'\i sica de Part\'\i culas}
\centerline{\it Universidad de Santiago}
\centerline{\it Santiago de Compostela, E-15706}
\vskip .9cm
\centerline{\bf Abstract}
\vskip .9cm
\noindent{$n$-Dimensional pure gravity theory can be obtained as the effective
theory of an $n+1$ model (with non-compact extra dimension) where general
$n+1$ reparametrization invariance is explicitly broken in the extra dimension.
As was pointed out in the literature, a necessary consistency condition for having 
a non-vanishing four dimensional
Newton constant is the normalizability in the extra dimension of the 
zero mass graviton. This, in turn, implies that gravity localization is produced
around the local minima of a potential in the extra dimension. 
We study gravity in the neighborhood of the soft (``thick") local minima.
}

\vskip2cm
\noindent US-FT-01/01
\Date{January, 2001}

\lref\rs{L. Randall and R. Sundrum, Phys.Rev.Lett. 83 (1999) 4690-4693, hep-th/9906064.}

\lref\h{C. Csaki, J. Erlich, T.J. Hollowood and Y. Shirman, 
Nucl.Phys. B581 (2000) 309-338, hep-th/0001033.}

\lref\exp{M.Bordag, B.Geyer, G.L. Klimchitskaya and V.M. Mostepanenko,
   Phys.Rev. D58 (1998) 075003, hep-ph/9804223.}

\lref\coop{F. Cooper, A. Khare and U. Sukhatme, Phys. Rep. 251 (1995) 267-385..}

\lref\dim{N. Arkani-Hamed, S. Dimopoulos and G. Dvali, Phys.Rev. D59 (1999) 086004, 
hep-ph/9807344.}

\lref\ga{J. Garriga and T. Tanaka, Phys.Rev.Lett. 84 (2000) 2778-2781, hep-th/9911055.}

\lref\gid{S.B. Giddings, E.Katz and L. Randall, JHEP 0003 (2000) 023, hep-th/0002091.}

\lref\cre{M. Gremm, Phys.Lett. B478 (2000) 434-438, hep-th/9912060.}

\lref\porr{G. Dvali, G. Gabadadze and M. Porrati, Phys.Lett. B484 (2000) 112-118, 
hep-th/0002190.}

\lref\dam{H. van Damm and M. Veltman, Nuc. Phys. B22 (1970) 397; V. I. Zakharov,
  JETP Lett. 12 (1970) 312.}

\lref\Porrb{M. Porrati, hep-th/0011152.}

\lref\Kogan{I.I. Kogan, S. Mouslopoulos and A. Papazoglou, hep-th/0011138.}

\lref\higu{A. Higuchi, Nucl. Phys. B282 (1987) 397; B325 (1989) 745.}

\lref\nuevo{C.D. Hoyle, U. Schmidt, B.R. Heckel, J.H. Gundlach, D.J. Kapner and
H.E. Swanson, hep-ph/0011014.}

\baselineskip 12pt

\newsec{Introduction}

In \refs{\rs} (that we will refer to as RSII model) 
it was shown that five dimensional gravity coupled to
a three brane (the ``brane-world") produces a geometry confining the zero mass graviton 
in a small neighborhood of the brane-world, in such a way that the propagation of
the massless gravitons are suppressed on the extra dimension. Moreover, 
the coupling to the brane-world of the continuum spectrum of massive gravitons 
propagating on the extra dimension was shown to be highly suppressed as well
(in the neighborhood of the brane-world).
Therefore, the massless gravitons propagate basically just along the brane-world
and the massive ones decouple from the brane-world physics.
This picture produces a four dimensional effective theory which is the one that we
observe (in the spirit of these models) at low energies.

The localization of the massless gravitons and the decoupling of the massive ones
were produced by the fact that these gravitons are under the influence
of a $\delta$-potential on the extra dimension, a consequence of the 
special geometry of the model (two copies of AdS-5 pasted along the
four dimensional brane-world). The physical interpretation of
such a sharp potential becomes from the fact that all the matter 
was just placed along the four-brane, breaking five-dimensional
general invariance in five dimensions with a $\delta (z)$ source.


In \refs{\h} the authors studied some properties that general $(n+1)$-dimensional
geometries have to fulfill in order to produce confinement of the 
massless gravitons in the extra dimension, and produce effective
theories in one dimension less. One necessary condition was found to be
that the massless graviton has to correspond to a normalizable wave
on the extra dimension. This corresponds to situations in
which the potential to with the gravitons are influenced in the extra dimension
has one (or several) minimum. The core of the massless gravitons is
placed around the minimum and then, physics can be effectively described as 
four-dimensional around such minimum, if the massive gravitons decouple. 

The RSII picture in which all matter is placed along a four dimensional 
submanifold of a five dimensional space-time is very suggestive from the string
theory point of view, where there exist objects like the $D-3$ branes
which can play the role of brane-worlds. This picture lead to metrics
which are discontinuous (with a jump on the brane) and, therefore, to singular
geometries.
However, to obtain localized gravity, the presence of a brane 
is not a must, and we will study a model where the metric is
continuous and smooth at the minima.
This restores the derivability condition on the metric and removes
the curvature $\delta$-singularity.

In our set up we will have several normalizable graviton modes, 
with a mass gap between them. In this respect, localized gravity
resembles a ``string" model. In fact, for the harmonic approximation, 
we will see that the mass spectrum
looks very close to that corresponding to a 
string with one excitation mode and with tension $\alpha '\sim {1\over w}$ 
(being $w$ the
frequency of oscillation). The lighter gravitons in our model should
describe a ``first order" local approximation to the lighter gravitons
of localized gravity with smooth-minima potentials. Of course,
our massive gravitons will differ substantially from the ``true"
massive ones at energy scales where the ``true" potential sensibly differs
from the harmonic one. However, the lightest gravitons are the most
sensible to gravitational tests.

We will work out a pure gravity case, in the sense that the dynamical fields
in five (and four) dimensions are just gravitons. The restriction to five
dimensions is not essential in our analysis, but we will take it as the
simplest one that produces an effective four-dimensional pure gravity.
The model that we will work out 
is based on the five dimensional action:

\eqn\Sc{
S=-{1\over {G_N^{(5)}}}\int{d^4 xdz\sqrt{-G} [R+\Lambda (z)+{\mu (z)\over{\sqrt{G_{zz}}}}]}.
}
$\Lambda (z)$ and $\mu (z)$ are source terms on the bulk and the $\mu (z)$ term
breaks, explicitly, five-dimensional general reparametrization invariance down to
four-dimensional reparametrization invariance along the $z$-direction 
(the RSII model corresponds to $\Lambda (z)$ constant and
$\mu (z)=\lambda\delta (z)$). The action \Sc\ can be interpreted as a 
continuous collection of parallel $D-3$ branes with local tension $\mu (z)$.
However, the important point is that the effective four dimensional
action derived from this action is just the Einstein-Hilbert action 
with a cosmological constant term (as we will see bellow).

The organization of the paper is as follows. In section 2 we present the
general setting of the models that we are going to study. 
We comment 
some general features of localized gravity on soft potentials at high and low
four-dimensional energies and also we briefly study experimental bounds
to the first massive graviton. 
In section 3 we
analyze localization of gravity by a harmonic confining potential. This
is a very interesting model, because it is analytically tractable.
We study the decoupling of heavier gravitons due to two facts, first the 
suppressed coupling
of the heavier gravitons to the ``thick world-brane" four-manifold, and second, 
an exponential suppression of the graviton waves along the extra dimension. 
We also study the form of the graviton waves in the extra dimension.
Finally, in section 4 we present our conclusions and give an outlook
of future developments.

\newsec{General Setting.}

First of all, let us discuss the effective four dimensional theory
associated to the action \Sc . For the ansatz:
\eqn\uno{
ds^2 = e^{2A(z)} (g_{mn}(x) dx^m dx^n +dz^2)
}
(where $z$ denotes the extra dimension) one gets, from \Sc , the equations of motion:
\eqn\em{
\eqalign{
&R^{(4)}_{mn}(x)+3 (A'' (z)+A' (z)^2)g_{mn}(x)={1\over 2} (e^{2A(z)}\Lambda (z)+
e^{A(z)}\mu (z))g_{mn}(x)\cr
&6 A' (z)^2-{1\over 2} R^{(4)}(x)={1\over 2} e^{2A(z)} \Lambda (z)\cr
}}
(the RSII case \refs{\rs} is the particular solution $g_{mn}=\eta_{mn}$, 
$A(z)=-Ln(1+k|z|)$,
$\Lambda (z)=12k^2$ and $\mu (z)=-12 k \delta (z)$).

Note that the second of these equations imply that 
$g_{mn}(x)$ must be a metric of constant scalar curvature $R^{(4)}(x)=k$. Therefore,
one gets the four dimensional effective action:
\eqn\eff{
S_{eff}=-{1\over {G_N^{(4)}}}\int{d^4x \sqrt{-g}R^{(4)}(x)}+\Lambda^{(4)}
\int{d^4x \sqrt{-g}}
}
being $R^{(4)}(x)$ the scalar curvature associated to the 
four dimensional metric $g_{mn}(x)$ and the effective four-dimensional Newton and
cosmological constants given by:
\eqn\cons{
\eqalign{
&G_N^{(4)}={G_N^{(5)}\over{\int_{-\infty}^{\infty}{dz e^{3A(z)}}}}\cr
&\Lambda^4 =-{2\over 3G_N^{(5)}} \int_{-\infty}^{\infty}{dz {d^2 e^{3A(z)}\over
{dz^2}}}+{k\over {2G_N^{(5)}}}\int_{-\infty}^{\infty}{dze^{3A(z)}}.\cr
}}

Demanding the four dimensional effective Newton 
constant $G_N^{(4)}$ to be non-vanishing requires that the function 
$e^{3A(z)}$ belongs to the Hilbert space $L^2[R]$ (\refs{\h}). Then, the boundary
contribution to $\Lambda^{(4)}$ in \cons\ vanishes identically.
Here, we will study the case where $e^{3A(z)}$ has just one
maximum (which will correspond to the case in which the confining
gravitational potential has just one minimum),
because we are interested on the local properties of gravity in the neighborhood of
a single minimum. Performing a global
coordinate translation in the $z$ direction and a global rescaling,
we can set this maximum at $z=0$ and fix $A(0)=0$. Then, we get:
\eqn\consb{
\eqalign{
&G_N^{(4)}=G_N^{(5)} \psi_0 (0)^2\cr
&\Lambda^{(4)}={k\over{2 G_N^{(4)}}}\cr
}}
where we have defined $\psi_0 (z)$ as the {\it normalized} function of
$L^2[R]$:
\eqn\dos{
\psi_0 (z)\equiv {e^{{3\over 2} A(z)}\over{\sqrt{\int_{-\infty}^{\infty}{dz e^{3A(z)}}}}}.
}

In this paper, we concentrate on effective four-dimensional theories where
$R^{(4)}(x)=0$, therefore we will take, from now on, $k=0$ and thus, the four dimensional 
effective cosmological constant identically vanishes.

Let us now study the linear metric perturbations that a gravitational 
source $j(x',z')$, in five dimensions, produces
around the four-dimensional flat vacuum solution. On the gauge $\gamma_{zz}=
\gamma_{nz}=0$ we can take, as the general perturbed metric: 
\eqn\tres{
ds^2=(e^{2A(z)}\eta_{mn}+\gamma_{mn}(x,z))dx^m dx^n+e^{2A(z)}dz^2.
}
where $\eta_{nm}$ is the four-dimensional Minkowski metric and
we are using the $(-,+,+,+)$ signature convention.
Imposing, moreover, the traceless transverse 
conditions $\gamma^n_n=\partial_n \gamma^n_m=0$ (where the 
contractions are performed with $\eta^{nm}$) and using standard techniques
(\refs{\ga,\h,\gid}) we arrive at:
\eqn\nuev{
\gamma (x,z)=e^{2A(z)}\int{d^4 x' dz' \Delta (x,z;x',z')j(x',z')}
}
where we suppressed the
Lorentz indices in the previous equation (corresponding to the graviton
polarizations) due to the decoupling of the equations on the chosen gauge.
$\Delta (x,z;x',z')$ is the retarded propagator ($t>t'$) given by:
\eqn\ocho{
\Delta (x,z;x',z')=e^{-{3\over 2}(A(z)-A(z'))}\sum_{n=0}^{\infty}\psi_n (z)\psi_n (z')
T_n (x,x')
}
where $\psi_n (z)$ are the {\it normalized} eigenfunctions of the operator
\eqn\O{
{\cal O}_z [A]\equiv {d^2\over{dz^2}}-{3\over 2} A''(z)-({3\over 2})^2 A'(z)^2
}
with eigenvalues $-m^2_n$:
\eqn\nueve{
{\cal O}_z [A]\psi_n (z)=-m^2_n\psi_n (z)
}
(in such a way that the sum in \ocho\ has to be understood as an 
integration on the continuum part of the spectrum)
and $T_n (x,x')$ is the standard 
four dimensional scalar retarded propagator for a particle of mass
$m_n$:
\eqn\diez{
T_n (x,x')=-{\theta (t-t')\over{2\pi}}(\delta [-(x-x')^2]-
  {m_n\over 2} {\theta [-(x-x')^2]\over\sqrt{-(x-x')^2}}J_1 [m_n\sqrt{-(x-x')^2}]).
}
(Here, $\theta$ is the step function $\theta (x)=0$ for $x\leq 0$ and $\theta (x)=1$
for $x>0$,
$-(x-x')^2=(t-t')^2-({\bar x}-{\bar x}')^2$ and $J_1(x)$ is the first Bessel 
function). The first term in $T_n(x,x')$ is the massless contribution
with support on the four-dimensional light-cone and the second one is
the contribution of the massive modes with support inside the light-cone.

${\cal O}_z[A]$ is an Hermitean operator that can be written as (\refs{\h}) 
\eqn\W{
{\cal O}_z[A]=
-W_z[A] W_z^{\dagger}[A]
} 
being 
\eqn\once{
W_z [A]=-{d\over{dz}}-{3\over 2}A'(z).
} 
Therefore, one gets that
the spectrum of ${\cal O}_z [A]$ is negative definite, in such a way that
$m^2_n\ge 0$ in \nueve . Also, the normalized massless zero mode is 
the normalized solution of the equation $W_z^{\dagger}[A]f(z)=0$,
which can be easily integrated to give \dos .

One can compute the static retarded propagator $\Delta ({\bar x},z;{\bar x}',z')$
from equations \ocho\ and \diez . The result is:
\eqn\doce{
\Delta ({\bar x},z;{\bar x}',z')\equiv\int_{-\infty}^{\infty}dt' \Delta (x,z;x',z')=
-{e^{-{3\over 2}(A(z)-A(z'))}\over{4\pi r}}\sum_{n=0}^{\infty}
e^{-m_n r}\psi_n (z)\psi_n (z').
}
(here, $r=|{\bar x}-{\bar x}'|$), where we get the standard Yukawa couplings.
This lead to the static metric perturbation (produced by a static source)
\eqn\trece{
\gamma ({\bar x},z)=-{1\over{4\pi}} e^{{1\over 2}A(z)}
\int_{-\infty}^{\infty}{d{\bar x}'dz' {e^{{3\over 2}A(z')}\over r}\sum_{n=0}^{\infty}
e^{-m_n r}\psi_n (z)\psi_n (z') j({\bar x}',z')}.
}

From the previous expression it is clear that, at low energy experiments
(large $r$), the lighter gravitons are the relevant ones. Also
we see the physical relevance of the four dimensional
submanifold $z=0$ (where we have set the maximum of the function $e^{A(z)}$).
First, the $j({\bar x}',z')$ part with support off the $z=0$ plane is suppressed 
by the factor $e^{{3\over 2}A(z')}$. Second, the propagation of gravitational
perturbations are suppressed off that plane by a factor $e^{{1\over 2} A(z)}$ as well.
In this sense one says that gravity is localized around the $z=0$ plane 
and, by abuse of language, we call this plane a ``thick world-brane".

Note also that  the linear
approximation will remain valid even if we approach the high energy 
four-dimensional limit ($r$ small)
for strong enough decreasing functions $e^{A(z)}$. The closer we go to the $r\to 0$ limit,
the farther we have to go in $z$ to maintain the linear approximation
valid ($z\to\infty$). In other words, the fifth dimension becomes more apparent as we perform
higher four dimensional experiments maintaining the graviton (i.e., linear)
approximation in our description of the physical processes.

Let us compute now the post-Newtonian law that an observer in five dimensions
perceives. Introducing the local perturbation $j({\bar x'},z')=
G_N^{(5)}M\delta^{(3)}({\bar x}')\delta(z')$ on equation \trece\ one gets:
\eqn\catorce{
\gamma ({\bar x},z)=-{G_N^{(5)}M\over{4\pi r}}e^{{A(z)\over 2}}
\sum_{n=0}^{\infty}e^{-m_n r}\psi_n (z)\psi_n (0).
}
(where now $r=|{\bar x}|$). In particular, for a four dimensional observer living 
in the $z=0$ four-manifold:
\eqn\quince{
\gamma ({\bar x},0)=-{G_N^{(4)}M\over{4\pi r}}(1+{1\over{\psi_0 (0)^2}}
\sum_{n=1}^{\infty} e^{-m_n r}\psi_n (0)^2).
}
(where we have used \consb ).

The previous equation resembles a ``mass renormalization"
(of a four dimensional Quantum Field Theory)\foot{We
find a {\it formal} analogy with the renormalized
Coulomb law in four dimensional QED.}, being
the ``bare" mass $M$ and the ``renormalized" 
(radius dependent) mass $M^{*}(A,r)$ defined through:
\eqn\dseis{
M^{*}(A,r)=M(1+{1\over{\psi_0 (0)^2}}\sum_{n=1}^{\infty} e^{-m_n r}\psi_n (0)^2).
}
(the dependence of $M^{*}$ on $A$ is due to the dependence on the spectrum of the operator
${\cal O}_z [A]$ in \O ). Note that, very far from the source, $\lim_{r\to\infty}
M^{*}(A,r)\to M$, as the only gravitons that can reach such distances are the 
massless ones. This result does not depend on the function $A(z)$, which
means that the details of the geometry in the fifth dimension is becoming
fuzzy to asymptotic four-dimensional observers.
On the other hand, on the ultrashort distances
$\lim_{r\to 0} M^{*}(A,r)\to {1\over{\psi_0 (0)^2}} M\delta (0)$ 
(where we have used the {\it formal}
identity $\sum_1^{\infty}\psi_n (0)^2\,$``$=$"$\,\delta (0)-\psi_0 (0)^2$).
Therefore, the mass corrections are becoming stronger at
shorter distances for observers on the ``world brane".
Note, however, that the linear approximation to $\gamma ({\bar x},0)$
becomes unjustified at distances of the order of the 
``dressed horizon" $G_N^{(4)}M^{*}(A,r_0)\sim r_0$, 
as $\gamma ({\bar x},0)$ in \quince\ is not
small. (In fact, the linear approximation to
the pure post-Newtonian law (discarding the massive modes) becomes also
unfair at distances shorter that the order of the ``bare horizon" $r\sim G_N^{(4)} M$).

We already commented above that linear approximations at 
ultrashort distances can be only justified if $A(z)$ is strong enough
and {\it always} at large $z$ distances (i.e., in five dimensions). 
Now we can make this statement
more precise. Let us go beyond the ``horizon" (i.e., to the
$r<G_N^{(4)} M$ region) maintaining the linear
approximation valid (i.e., in five-dimensional space). From \catorce\ we have that
\eqn\ccatorce{
{G_N^{(4)}M\over r} e^{{A(z)\over 2}}\sim 1\,\,\,\,\Rightarrow 
A(z)\sim 2 Ln\Bigl( {r\over{G_N^{(4)} M}}\Bigr)
}
therefore, the linear approximation is valid if $A(z)$ is 
smaller than $\sim 2 Ln\Bigl( {r\over{G_N^{(4)} M}}\Bigr)$. Being 
$e^{A(z)}$ normalizable, this means big $z$.

Let us say a word on experimental bounds to the Yukawa-like corrections to
the post-Newtonian law. From \quince\ it is clear that the first Yukawa correction
to the post-Newtonian law is given by the first massive graviton (which
we call $\psi_{*}(z)$)   which couples to the manifold $z=0$ (i.e., 
such that $\psi_{*}(0)\ne 0$). Moreover, this Yukawa correction
is certainly the most important one if the higher mass modes couple less to $z=0$
than $\psi_{*}(z)$ (i.e., if $\psi_{*}(0)^2 >\psi_{n}(0)^2$ $\forall\,\, n>*$).
This must be the case for localized gravity.

In reference \refs{\exp} , experimental bounds for Yukawa-like corrections
to the gravity potential of the form:
\eqn\yuk{
V(r)=-\alpha N_1 N_2 \hbar c {e^{-{r\over\lambda}}\over r}
}
are studied for a Casimir-force experiment. $\alpha$ is a dimensionless
parameter, $N_1$ and $N_2$ are the numbers of nucleons in the atomic
nuclei which are interacting gravitationally in the experimental setting,
and $\lambda$ is a 
parameter with dimensions of length (in meter units). 
The experimental bound is a curve
on the $(\alpha ,\lambda)$-plane. The regions of 
the $(\alpha ,\lambda)$-plane below that curve are permitted, and above
are ruled out by experiments (see figure $3$ of \refs{\exp}).

In our present case, we have to compare \yuk\ with the first Yukawa correction
that the post-Newtonian law \quince\ gives for the gravitational interaction 
between two nuclei with $N_1$ and $N_2$ nucleons:

\eqn\yukd{
v(r)=-{G_{N}^{(4)} m_{N}^2 N_1 N_2 \over{4\pi r}} {\psi_{*}(0)^2\over{\psi_{0}(0)^2}}
e^{-m_{*} r}
}
(note that the corrections of the higher mass modes are smaller due both
to the exponential mass suppression and the $z=0$ coupling suppression).
For our case, then, $\alpha$ and $\lambda$ are given by: 
\eqn\al{
\alpha\sim 5\cdot 10^{-40} {\psi_{*}(0)^2\over{\psi_{0}(0)^2}}, \,\,\,\,\,\,\,\,\,\,\, 
\lambda={1\over {m_{*}}}.
}
Because we are assuming that gravity is localized,
${\psi_{*}(0)^2\over{\psi_{0}(0)^2}}<1$, then we
have that $\alpha <5\cdot 10^{-40}$. Therefore this shows that this scenario
can not be tested by the done pure gravity experiments
(Casimir, Cavendish and Eotvos-type, which reach values over
$\alpha\sim 10^{-39}$) in \refs{\exp}.

On the other hand, very recently, a different experiment has been studied in 
\refs{\nuevo} which is able to set a lower bound on $m_{*}$. In their notation, 
the Yukawa correction
to the gravitational potential is parametrized by
\eqn\cuno{
V(r)=-G_N^{(4)}{m_1 m_2\over r} (1+\alpha e^{-{r\over\lambda}}).
} 
In our case, and from figure 4 of \refs{\nuevo}, we then get:
\eqn\cdos{
\alpha =\bigl( {\psi_{*} (0)\over {\psi_{0} (0)}}\bigr)^2<1 \,\,\,\Rightarrow
\lambda_{*}<2,5\cdot 10^{-4} met.\,\,\, \Rightarrow m_{*}>7.5\cdot 10^{-4} eV.
}
Note that pure gravitational test puts a very low lower bound
on the mass of the lightest graviton that couples to the ``thick world-brane".

\newsec{The Harmonic Oscillator Approximation.}

Demanding that the zero mode of the spectrum associated to the linear operator 
${\cal O}_z [A]$ in \O\ is normalizable, translates to the fact that
the ``Schrodinger"-like potential in equation \nueve\ has (at least) a minimum.
The extreme case in which the only normalizable mode is the massless 
mode corresponds to the RSII model, where the $\delta (z)$-potential
can locate just one bound state. Certainly, from a conventional General Relativity
point of view, it is a more pleasant picture
to have a potential that is not that sharp, and that it is smooth also at the
minimum. This lead us to trade 
(see also \refs{\cre, \h})
the $\delta (z)$ potential by a smooth and derivable potential. This,
in turn, maintains (apart of continuity) a derivability property on the 
five-dimensional geometry (in such a way that we do not have to 
worry about Israel matching conditions as in the AdS cases).
At this point, we can approximate (locally) the gravitational
confining potential around its minima in the $z$ direction by 
the harmonic oscillator potential. The energies up to which this approximation
is faithful depend on the energy scale at which the ``true" potential 
differs significantly from the harmonic one. But at least, locally, 
the massless graviton should be accurately described
by the ground state of the harmonic oscillator (and perhaps, some
of the first massive gravitons). Remember that, precisely, the smaller
the masses of the gravitons are, the longer they can travel to large 
``world-brane"
distances (see \quince ), where we measure the deviations to
the Newtonian law. 
Therefore, the light gravitons are the most relevant ones for
experimental purposes. 

Let us mention at this point that the exact spectral resolution of the problem 
\nueve\ is known
for some few potentials. In particular, we have the ``Shape Invariant Potentials" 
of supersymmetric quantum mechanics \refs{\coop} (one of those is the 
harmonic case). But we do not expect them to change significantly the local
picture (at least for the light gravitons) around the soft minima of the potential.

Therefore, let us take 
\eqn\dsiete{
A(z)=-{w\over 3} z^2
}
in equation \nueve . The spectral resolution to this equation is
given by:
\eqn\docho{
\psi_n (z) =({w\over\pi})^{1\over 4} {1\over{\sqrt{2^n n!}}} e^{-{w\over 2}z^2}
 H_n(\sqrt{w}z),\,\,\,\,\,\,\,\,\,\,m_n=\sqrt{2nw}, 
}
where $H_n(x)$ are Hermite polynomials. A few comments are in order. First
of all, from equations \consb\ and \docho , 
the harmonic frequency of vibration can be expressed
in terms of the ratio of the Newton constants $G_N^{(4)}$ and
$G_N^{(5)}$ as:
\eqn\dnueve{
{G_N^{(4)}\over{G_N^{(5)}}}=\sqrt{w\over\pi}.
}

Second, the mass spectrum \docho\ is very similar to the mass spectrum
of a string with one excitation mode and tension
\eqn\str{
m_{str}^2\sim {n\over {\alpha '}}\,\,\,\,\,\Rightarrow\,\,\,\,\, \alpha '\sim {1\over w}.
}
In fact, the graviton wave-modes along the extra dimension can
be interpreted as the vibrating modes of a ``string", and the mass associated
to a given mode is proportional to its number of oscillation.
A difference with string-theory is that, there, the strings are
of finite length and one imposes Newmann or Dirichlet 
boundary conditions. In our present case, the wave satisfy the  
(``finite energy") boundary conditions on the extra dimension 
(see \nuev\ and \ocho ) $\lim_{z\to\pm\infty} \gamma (x,z)=0$
which are analogous to a string configuration with
two D3-branes at $z\to\pm\infty$ with a string attached to them. 

Also, from \docho , the mass spectrum 
of the gravitons tends to 
be closer as one increases the energy 
$m_{n+1}-m_n\sim\sqrt{{w\over{2n}}}$.
Finally, the odd modes of the gravitons ($n$ odd) do not couple to
the four dimensional brane-world at $z=0$ (just because $H_{2n+1}(0)=0$).
Therefore, the highest correction to the Newton law comes from the
$n=2$ graviton. 

Let us then discuss the experimental bound to the 
frequency of vibration of the harmonic oscillator (i.e., to the ratio 
of the four dimensional effective Newton constant and 
the five-dimensional Newton constant). 
Because, for the harmonic case, $\alpha =1/2$ (in the notation of 
\refs{\nuevo}), the mass bound for the lightest massive graviton
does not change substantially the result in \cdos . In particular,
one gets $m_2>5\cdot 10^{-4}$ eV, $w> 6\cdot 10^{-8}$ ${\rm eV}^2$ and
$G_N^{(4)}>1.4\cdot 10^{-4} G_N^{(5)}$ eV.
Note that this mass bound is very small compared with the electron mass!,
and yet, it is compatible with done pure gravity experiments. 

In spite of that we are working with an uncompactified extra dimension, 
from \consb\ one can define $R\equiv \psi (0)^{-2}$ as the 
{\it analog} of ``Kaluza-Klein compactification radius". Then, from \docho\
and the experimental lower bound for $w$ that we got, we 
obtain an experimental upper bound for $R$:
\eqn\R{
R<9\cdot 10^{-4} {\rm m}\sim 1 {\rm mm}
}
which is concordant with the results in the literature in 
which the pure gravity experiments are compatible with
extra dimensions with a compactification scale lower than
$1$ mm. 

Note that a ``unification" five-dimensional Newton constant (with a 
five-dimensional Planck mass of the order of a TeV) is deep inside
the experimentally ruled-out region, 
because ${G_{N}^{(4)}\over {G^{(5)GUT}_N}}\sim 10^{-21}$ eV. This is  
concordant also with the unification models of \refs{\dim}, which require
two or more extra dimensions.

Let us now discuss the contribution of the massive modes to the
post-Newtonian law. From equation \docho\ we get
that the odd modes decouple completely from $z=0$, and the coupling
of the even modes to $z=0$ are driven by (see \dseis )
\eqn\veinte{
c_{2n} (r)\equiv e^{-m_{2n} r}\bigl( {\psi_{2n}(0)\over {\psi_0 (0)}}\bigr
)^2\equiv e^{-m_{2n} r}\lambda_{2n}.
}
Using the doubling formula for the Euler Gamma function, we arrive at:
\eqn\vuno{
\lambda_{2n}={1\over{\sqrt{\pi}}} {\Gamma (n+{1\over 2})\over{\Gamma (n+1)}}.
}
It is clear that there are two suppressing factors for the contribution
to the post-Newtonian law when the graviton mass increases. The first one
($\lambda_{2n}$) is due to the fact that the coupling 
(to the ``world-brane" at $z=0$) of the eigenfunctions of the 
spectral problem \nueve\ decreases 
as we increase $n$. The second one is due to the exponential suppression
with the ``world-brane" distance $r$ along the $z=0$ plane 
$e^{-m_{2n} r}$. This two suppression factors is what 
we have drawn in figure $1$, for the first lighter fifty 
gravitons: $\lambda_{2n}$ (in gray) and 
$c_{2n} (r)$ for $r=1$mm (in black), and for the most 
sensitive value of $w$ compatible with the experimental data 
in \refs{\nuevo} ($2\sqrt{w} \sim 10^3 {\rm m}^{-1}$). At $r=1$cm
$c_{2n}(r)$ is a number smaller than $10^{-8}$ for the 
gravitons with $n\geq 1$, therefore, at these energies only the massless
graviton couples to ``brane-world" physics.

Due to the discrete character of the graviton spectrum, we do not face
problems with the van Dam-Veltman-Zakharov discontinuity (\refs{\dam,
\higu, \Porrb ,\Kogan}). In our case
the contribution of the first massive-mode to the bending of
light by the sun has a suppression factor $e^{-2\sqrt{w} r}$,
that, for the lower bound on $w$ ($2\sqrt{w} \sim 10^3 {\rm m}^{-1}$) and
distances bigger than the sun's radius, is completely
negligible.


\vskip 0.5cm
\centerline{\hskip.2in\epsffile{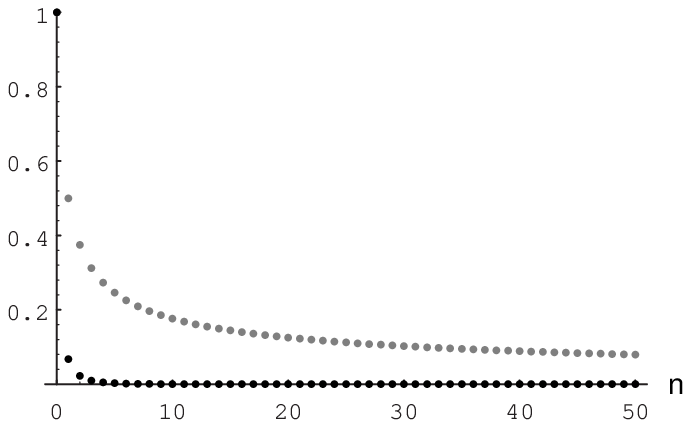}}
\vskip 0.3cm
\noindent\centerline{{\tenrm Fig. 1: In gray, $\lambda_{2n}$ for the first
fifty gravitons. In black, $c_{2n}(r)$ for $r=1mm$ and}} 
\noindent\centerline{{\tenrm $2\sqrt{w}=10^3 {\rm m}^{-1}$. For $r=1$cm,
$c_{2n}(r)$ is negligible for $n\geq 1$.
}}
\vskip 1cm

Let us now analyze the form of the graviton waves. In equation \catorce\
we observe that the $z$ dependence of the graviton modes 
corresponds to that of the eigenfunctions of \nueve\ {\it suppressed}
by an exponential factor $e^{{A(z)\over 2}}$. Moreover, the dependence
with the distance $r$ from the source along the $z=0$ plane
is given by the Yukawa factors times ${1\over r}$.
From equations \catorce ,
\dsiete\ and \docho\ we get:
\eqn\vdos{
\mu_{2n}\equiv
{1\over{r}}e^{{A(z)\over 2}}
\sum_{n=0}^{\infty}e^{-m_n r}\psi_n (z)\psi_n (0)=
(-)^n {e^{-2\sqrt{nw} r}\over{2^{2n}\Gamma (n+1)r}}e^{-{2\over 3}w z^2}
H_{2n}(\sqrt{w} z).
}
In figure 2 we draw $\mu_{2n}$ and compare the
graviton waves (for $r=1$mm and $r=1$cm and the same
mass source $M$) for the first three modes that couples to the
four-manifold at $z=0$. We have taken, 
again, the experimental lower
bound for $w$ ($2\sqrt{w} \sim 10^3 {\rm m}^{-1}$).
Also, we have drawn in gray the corresponding
harmonic oscillator quantum wave-function (normalized such that its
value on $z=0$ coincides with the $r=1$mm graviton wave), in such a way
that we have an explicit picture of the $e^{A(z)\over 2}$ suppression
factor in \catorce . In this figure, the gravitons $2n=2,4$ at 1cm 
do not appear because their amplitudes are $10^{-8}$ and $10^{-12}$
(respectively) smaller than for the case of 1mm.

\vskip 0.5cm
\centerline{\hskip.2in\epsffile{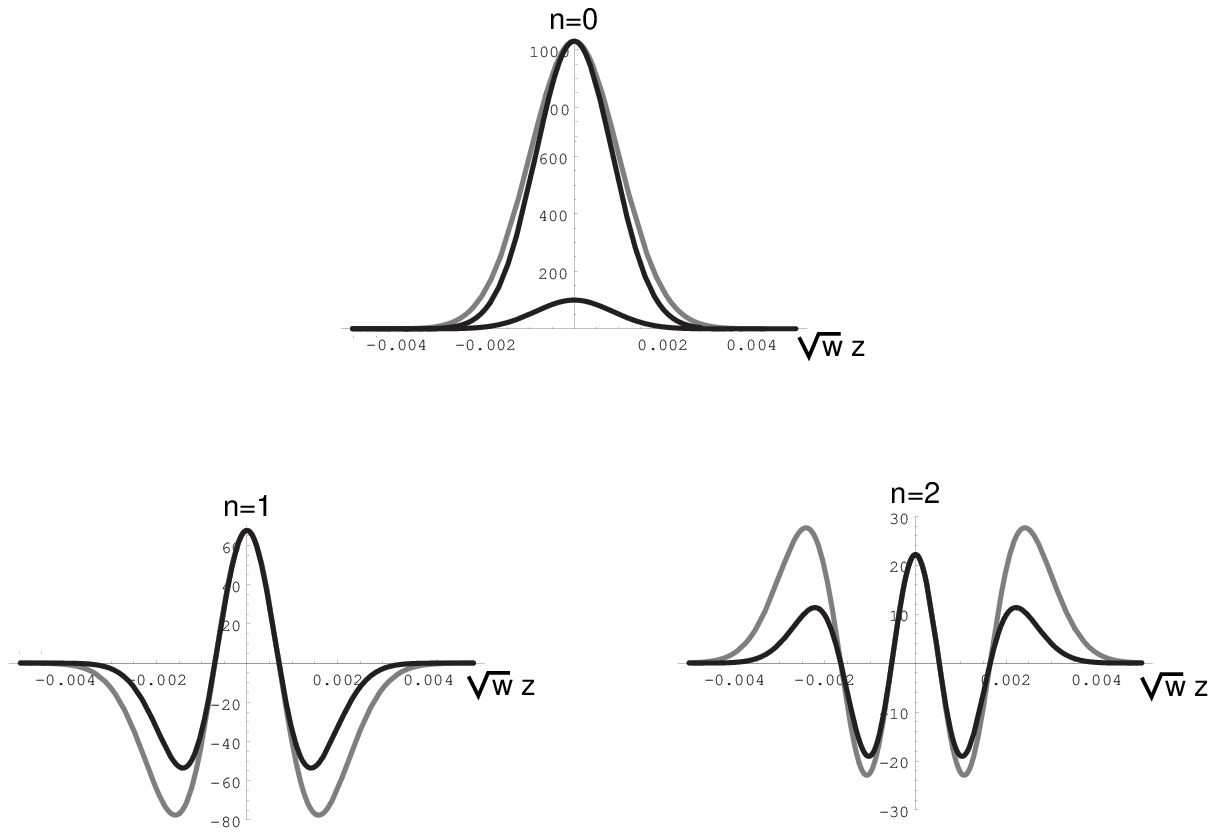}}
\vskip 0.3cm
\noindent\centerline{{\tenrm Fig. 2: In gray, quantum harmonic oscillator
wave-functions (properly normalized). In black,}}  
\noindent\centerline{{\tenrm 
$\mu_{2n}$ ($n=0,1,2$) for $r=1$mm (the highest ones) and  $r=1$cm (the lower ones).}} 
\noindent\centerline{{\tenrm For $n=1,2$
the amplitudes of the gravitons at $r=1$cm are negligible.
}}
\vskip 1cm

We can also naturally introduce a notion of the ``width" 
(at ``brane-world" distance $r$ from the source)
along the $z$ direction of the graviton waves 
$\gamma_{2n} ({\bar x},z)$ (the ones that couple to four dimensional physics)
through $l_{2n}(r)\equiv (<z^2>_{2n})^{1\over 3}$ being
\eqn\vtres{
<z^2>_{2n}\equiv \int_{-\infty}^{\infty}dz\,\, z^2 \gamma_{2n} ({\bar x},z)^2.
}
Substituting equations \catorce\ and \docho\ in \vtres\ we arrive at:
\eqn\vcuatro{
\eqalign{
<z^2>_{2n} &={1\over {w^{3\over2}}} \Big( {G_N^{(4)} M e^{-2\sqrt{n w}r}\over{4\pi r}}
{1\over{2^{2n} \Gamma (n+1)}}\Big)^2 \Gamma (2n+{1\over 2})({3\over 4})^{5\over 2}\times\cr
&\Bigl[ ({2\over 3}-8n)F[-2n,-2n;{1\over 2}-2n;2]-
      {64 n^2\over{4n-1}}F[1-2n,1-2n;{3\over 2}-2n;2]\Bigr]\cr
}}
being $F[\alpha,\beta;\gamma;\delta]$ Gauss's Hypergeometric series. 
As an example, from \vcuatro\ we can obtain the ratio of the lengths
corresponding to the massless graviton and the first two massive ones
which couples to the ``world-brane",
obtaining:
\eqn\vcinco{
{<z^2>_0\over {<z^2>_2}}={64\over 79}e^{4\sqrt{w}r}
\,\,\,\,\,\,\,
{<z^2>_0\over {<z^2>_4}}={16384\over 12753}e^{4\sqrt{2w}r}
}
We see that, for large $r$, the thickest graviton on the extra dimension
is the massless one .

In figure 3
we draw the $n$-dependent part of $<z^2>_{2n}$ in de range $0\leq n\leq 15$, 
for $r=1$mm and $2\sqrt{w} \sim 10^3 {\rm m}^{-1}$. Again, at $r=1cm$, 
the size of the gravitons with $n\geq 1$ is negligible for this
$w$. 

\vskip 0.5cm
\centerline{\hskip.2in\epsffile{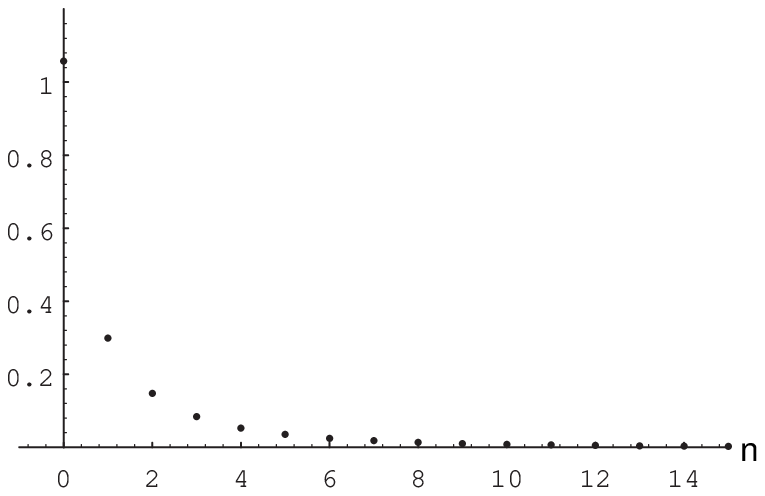}}
\vskip 0.3cm
\noindent\centerline{{\tenrm Fig. 3: n-dependent part of $<z^2>_{2n}$ for the first fifteen 
lighter gravitons.
}}
\vskip 1cm

\newsec{Conclusions and Outlook.}

We have studied general aspects of localized gravity of pure
gravity ``thick-branes". In the literature, it has been suggested
that the RSII-type models can be observationally best tested
in pure gravity experiments.
We have obtained an experimental
lower bound to the mass of the graviton which produces the biggest contribution
to deviations to the Newtonian law. We have
studied the local harmonic approximation to pure gravity localization (for four-dimensional
geometries of zero scalar curvature) and obtained the expression \dnueve\ for the
frequency of oscillation in terms of the ratio of the four-dimensional
effective Newton constant and the five-dimensional Newton constant. 
As we have argued, the harmonic approximation for the zero mode
(and maybe, some of the lighter ones) is a 
``first-order" description of the ``exact" first
graviton modes for ``thick-branes" (those that can allocate several
normalizable modes), in the neighborhood of the minima of the confining
gravitational potential.
We have studied for this
approximation (and analytically) the increasing relative 
decoupling of the massive modes to four dimensional physics and
the form of the graviton waves.
We have also introduced a notion of
size for the graviton waves on the extra dimension and found its analytical
expression. The heavier, the smaller they are in the 
extra dimension (due to the two exponential suppression factors
in \vdos ). 

There are several issues that remain for further research. 
One of them is to study ``localized thermodynamics" by
taking a graviton thermal-bath configuration as a source 
in \trece\ instead of a $\delta (z)$ source (profiting of the fact
that the harmonic case is analytically tractable). This means to impose
that the physics on the extra dimension is a quantum one,
in contrast with the classical one that we have studied
($\delta (z)=\sum_{n=0}^{\infty} \psi_n (z)\psi_n (0)$ are
not density matrix entries). The extra dimension model could
provide us some hints on black-hole thermodynamics and 
information problem.

Also, it would be interesting to introduce dynamics on the problem.
This can be done introducing time-dependent sources in \nuev ,
studying time-dependent background metrics ($A(z)\to A(z,t)$)
or studying transition amplitudes between the graviton states
by using perturbation theory up to order $\gamma (x,z)^2$. 

Another question is the implications of the picture presented
here for cosmology and the cosmological constant. 
We have seen that, for the models described by the action \Sc ,
the cosmological constant identically vanishes for 
four-geometries of vanishing scalar curvature (it is a surface term).
In the past years, experimental evidence for non-vanishing
four-dimensional cosmological constant has emerged.
Therefore, the study of the cases
in which the four-geometry in \uno\ is of constant (but non-vanishing)
scalar curvature (being the four-dimensional effective cosmological constant
in \consb\ not vanishing) is phenomenologically relevant. 
Moreover, the localized models that we have explored contain
massive modes for the gravitons,  whose effects on four
dimensional physics are highly suppressed already at decimeter scales
(by experimental bounds).
Therefore, the Newtonian law does not change 
at astronomic scales, but we could have cosmological implications (as this
massive gravitons mimic ``dark matter"). A cosmological analysis
could provide an experimental upper bound for $m_{*}$ and
refine the lower bound that we have found.

\bigskip
{\bf Acknowledgments}

We would like to thank Roberto Emparan, Jose M. S\'anchez de Santos and
Ricardo V\'azquez for valuable comments and for a critical reading 
of the manuscript. This work was supported in part by DGICYT under
grant PB96-0960 and by the Xunta de Galicia grant PGIDT00-PXI-20609.

\listrefs
\end